\documentclass[doublecol]{epl2}
\usepackage{amsfonts,amssymb,amsmath,color}
\usepackage{graphicx}
\usepackage{bm}

\begin{document}
\title{Testing time series {ir}reversibility using complex network methods}
\author{Jonathan F. Donges\inst{1,2,3} \and Reik V. Donner\inst{1} \and J\"urgen Kurths\inst{1,2,4}}
\shortauthor{J.F.~Donges, R.V.~Donner, and J. Kurths}
\institute{
	\inst{1}{Potsdam Institute for Climate Impact Research, P.O. Box 601203, 14412 Potsdam, Germany}\\
	\inst{2}{Department of Physics, Humboldt University Berlin, Newtonstr.~15, 12489 Berlin, Germany}\\
	\inst{3}{Stockholm Resilience Centre, Stockholm University, Kr\"aftriket 2B, 11419 Stockholm, Sweden}\\
	\inst{4}{Institute for Complex Systems and Mathematical Biology, University of Aberdeen, Aberdeen AB24 3FX, {UK}}
}


\date{\today}

\abstract{
The absence of time-reversal symmetry is a fundamental property of many nonlinear time series. Here, we propose a {new} set of statistical tests for time series {ir}reversibility based on standard and horizontal visibility graphs. Specifically, we statistically compare the distributions of time-directed variants of the common {complex network} measures degree and local clustering coefficient. Our approach does not {involve} surrogate data and {is applicable} to relatively short time series. We demonstrate its performance for paradigmatic model systems with known time-reversal properties as well as {for picking} up signatures of nonlinearity in neuro-physiological data.
}

\maketitle

\section{Introduction}

Nonlinear processes govern the dynamics of many real-world systems. Therefore, a sophisticated diagnostics and identification of such processes from observational data is a common problem in time series analysis important for model development. Consequently, in the last decades, testing for {non}linearity of time series has been of {great} interest. {Various} approaches {have} been developed for identifying signatures of different types of nonlinearity as a necessary precondition for the possible emergence of chaos~\cite[\S~5.3]{Tong1990}. 
Since linearity of Gaussian processes directly implies time-reversibility~\cite{Weiss1975,Lawrance1991,Diks1995} (see~\cite[\S~4]{Diks1999} for further details), {nonlinearity results (among other features) in an} asymmetry of certain statistical properties under time-reversal~\cite{Theiler1992}. Therefore, studying reversibility properties of time series is an important alternative to the direct quantitative assessment of nonlinearity~\cite{Voss1998}. {In contrast to classical higher-order statistics requiring surrogate data techniques~\cite{Theiler1992}, most recently developed approaches for testing irreversibility have been based on symbolic dynamics~\cite{Daw2000,Kennel2004,Cammarota2007} or statistical mechanics concepts~\cite{Costa2005,Porporato2007,Roldan2010}.}

Motivated by the enormous success of complex network theory in many fields of {science}~\cite{Newman2003}, in the last years several techniques for network-based time series analysis have been proposed~\cite{Zhang2006,Lacasa2008,Luque2009,Xu2008,Marwan2009,Donner2010NJP,Donner2011IJBC}. As a particularly successful example, \emph{visibility graphs (VGs)} and related methods~\cite{Lacasa2008,Luque2009} {(see Methods)} are based on the mutual visibility relationships {between points in a one-dimensional landscape representing a} univariate (scalar-valued) time series. The degree distributions of the thus constructed VGs allow classifying time series according to the type of recorded dynamics and obey characteristic scaling in case of fractal or multifractal behaviour of the data under study~\cite{Lacasa2009,Ni2009}. These relationships make VGs promising candidates for studying observational time series from various fields of research such as turbulence~\cite{Liu2010}, finance~\cite{Ni2009,Yang2009,Qian2010}, physiology~\cite{Lacasa2009,Ahmadlou2010}, or geosciences~\cite{Elsner2009,Tang2010,Donner2012AG,Telesca2012,Pierini2012}.

In~\cite{Lacasa2012}, Lacasa~\textit{et~al.} demonstrated that \textit{horizontal visibility graphs (HVGs)}~\cite{Luque2009}, an algorithmic variant of VGs (see Methods), allow discriminating between reversible and irreversible time series. Based on a time-directed version of HVGs, they could show that irreversible dynamics results in an asymmetry between the probability distributions of the numbers of incoming and outgoing edges of all network vertices, which can be detected by means of the associated Kullback-Leibler divergence. In this work, we {thoroughly} extend this idea and provide a set of rigorous statistical tests for time series {ir}reversibiliby, which can be formulated based on both standard and horizontal VGs {and utilise different network properties}. Specifically, we demonstrate that for VGs and HVGs, degrees as well as local clustering coefficients can be decomposed into contributions from past and future observations, which allows studying some of the time series' statistical properties under time-reversal. We find statistically significant deviations between the distributions of time-ordered vertex properties for nonlinear systems for which the absence of time-reversal symmetry is known, but not for linear systems. As a real-world application, the power of the proposed approach for studying the presence of nonlinearity in neuro-physiological time series (EEG recordings) is demonstrated.

\section{Methods}

Visibility graphs are based on a simple mapping from the time series to the network domain exploiting the local convexity of scalar-valued time series $\{x(t_i)\}_{i=1}^N$. Specifically, each observation ${x_i=}x(t_i)$ is assigned a vertex $i$ of a complex network, which is uniquely defined by the time of observation $t_i$. Two vertices $i$ and $j$ are linked by an edge $(i,j)$ iff the condition~\cite{Lacasa2008}
\begin{equation}
x_k<x_j+(x_i-x_j)\frac{t_j-t_k}{t_j-t_i}
\end{equation}
\noindent
applies for all vertices $k$ with $t_i<t_k<t_j$. This is, the adjacency matrix $(A_{ij})$ describing the VG as a simple undirected and unweighted network reads
\begin{equation}
A_{ij}^{({\text{VG}})}=A_{ji}^{({\text{VG}})}=\prod_{k=i+1}^{j-1} \Theta\left(x_j+(x_i-x_j)\frac{t_j-t_k}{t_j-t_i}-x_k\right),
\end{equation}
\noindent
where $\Theta(\cdot)$ is the Heaviside function. 

Horizontal VGs {provide} a simplified version of this algorithm~\cite{Luque2009}. For a given time series, the vertex sets of VG and HVG are the same, whereas the edge set of the HVG maps the mutual horizontal visibility of two observations $x_i$ and $x_j$, i.e., there is an edge $(i,j)$ iff $x_k<\min(x_i,x_j)$ for all $k$ with $t_i<t_k<t_j$, so that
\begin{equation}
A_{ij}^{({\text{HVG}})}=A_{ji}^{({\text{HVG}})}=\prod_{k=i+1}^{j-1} \Theta\left(x_i-x_k\right) \Theta\left(x_j-x_k\right).
\end{equation}
\noindent
VG and HVG capture essentially the same properties of the system under study (e.g., regarding fractal properties of a time series), since the HVG is a subgraph of the VG with the same vertex set, but possessing only a subset of the VG's edges. Note that the VG is invariant under a superposition of linear trends, whereas the HVG is not.

\subsection{Time-directed vertex properties} 

{Since the definition of VGs and HVGs takes the timing (or at least time-ordering) of observations explicitly into account,} the direction of time is intrinsically interwoven with the {resulting network} structure. {To account for this fact, we} define a set of novel statistical network quantifiers {based on} two simple vertex characteristics:

(i) On the one hand, the degree {$k_i =\sum_{j} A_{ij}$} measures the number of edges incident to {a given} vertex {$i$}. For a (H)VG, we {can} decompose this quantity for a vertex corresponding to a measurement at time $t_i$ into contributions due to other vertices in the past and future of $t_i$,
\begin{eqnarray}
k_i^r &= \sum_{j<i} A_{ij},\\
k_i^a &= \sum_{j>i} A_{ij}
\end{eqnarray}
{with $k_i=k_i^r+k_i^a$}, being referred to as the \emph{retarded} and \emph{advanced degrees}, respectively, in the following. Note that $k_i^r$ and $k_i^a$ correspond to the respective in- and out-degrees of time-directed (H)VGs as recently defined in~\cite{Lacasa2012}. {While the degrees of an individual vertex can be significantly biased due to the finite data~\cite{Donner2012AG}, the resulting frequency distributions of retarded and advanced degrees are equally affected. Since the method to be detailed below is exclusively based on these distributions, we will not further discuss this question here.}

(ii) On the other hand, the local clustering coefficient $\mathcal{C}_i = {k_i \choose 2}^{-1} \sum_{j,k} A_{ij} A_{jk} A_{ki}$ is another vertex property of higher order characterising the neighbourhood structure of vertex $i$~\cite{Newman2003}. Here, for studying the connectivity due to past and future observations separately, we define the \emph{retarded} {and} \emph{{advanced local clustering coefficients}}
\begin{eqnarray}
\mathcal{C}_i^r &= {k_i^r \choose 2}^{-1} \sum_{{j<i,k<i}} A_{ij} A_{jk} A_{ki},\\
\mathcal{C}_i^a &= {k_i^a \choose 2}^{-1} \sum_{{j>i,k>i}} A_{ij} A_{jk} A_{ki}.
\end{eqnarray}
Hence, both quantities measure the probability that two neighbours in the past (future) of observation $i$ are mutually visible themselves. Note that the decomposition of $\mathcal{C}_i$ into retarded and advanced contributions is not as simple as for the degree and {involves} degree-related weight factors and an additional term combining contributions from the past and future of a given vertex.

\subsection{Testing for time-{ir}reversibility}

Time-{ir}reversibility of a stationary stochastic process or time series $\{x_i\}$ requires that for arbitrary $n$ and $m$, the tuples $(x_n,x_{n+1},\dots,x_{n+m})$ and $(x_{n+m},x_{n+m-1},\dots,x_n)$ have the same joint probability distribution~\cite{Lawrance1991}. Instead of testing this condition explicitly (which is practically unfeasible in most situations due to the necessity of estimating high-dimensional probability distribution functions from a limited amount of data), {for detecting time series irreversibility it can be sufficient} to compare the distributions of certain statistical characteristics obtained from both vectors {(e.g.,~\cite{Tong1990})}. {Following the decomposition of vertex properties into time-directed contributions proposed above,} {(H)VG-based} methods appear particularly suited for this purpose. Specifically, in the following we will utilise the frequency distributions {$p(k^{r})$ and $p(k^{a})$ ($p(\mathcal{C}^{r})$ and $p(\mathcal{C}^{a})$)} of retarded and advanced {vertex properties} as representatives for the statistical properties of the time series when viewed forward and backward in time. 

In the case of time-reversibility, we conjecture that both sequences {$\{k_i^{r}\}$ and $\{k_i^{a}\}$} (or {$\{\mathcal{C}_i^{r}\}$ and $\{\mathcal{C}_i^{a}\}$}) should be drawn from the same probability distribution, because the visibility structure towards the past and future of each observation has to be statistically equivalent. In turn, for an irreversible (i.e., nonlinear) process, we expect to find statistically significant deviations between the probability distributions of retarded and advanced characteristics. 

As an alternative to the Kullback-Leibler distance between the empirically observed distribution functions used by Lacasa~\textit{et~al.}~\cite{Lacasa2012}, we propose utilising some standard statistics for testing the homogeneity of the distribution of random variables between two independent samples. In this framework, rejecting the null hypothesis that {$\{k_i^{r}\}$ and $\{k_i^{a}\}$} ({$\{\mathcal{C}_i^{r}\}$ and $\{\mathcal{C}_i^{a}\}$}) are drawn from the same probability distribution, respectively, is equivalent to rejecting the null hypothesis that the time series under investigation is reversible. {Since for sufficiently long time series (representing the typical dynamics of the system under study), the available samples of individual vertex properties approximate the underlying distributions sufficiently well, we can (despite existing correlations between subsequent values)} consider the Kolmogorov-Smirnov (KS) test {for} testing this null hypothesis. {Specifically}, a small $p$-value of the KS test statistic (e.g., $p<0.05$) implies that the time series has likely been generated by an irreversible stochastic process or dynamical system. Even more, the{se} $p$-values are distribution-free in the limit of $N\to\infty$. {Neglecting possible effects of the intrinsic correlations between the properties of subsequent vertices on the estimated $p$-values (which shall be addressed in future research), this} implies that we do \textit{not} need to construct surrogate time series for obtaining critical values of our test statistics as in other {ir}reversibility tests. {Note that other (not network-related) statistical properties sensitive to the time-ordering of observations could also be exploited for constructing similar statistical tests for time series irreversibility. A detailed discussion of such properties is, however, beyond the scope of this Letter.}

\section{Model systems}

Let us illustrate the potentials of the proposed method for two simple model systems: (a) a linear-stochastic first-order autoregressive (AR(1)) process
\begin{equation}
x_t=\alpha x_{t-1}+\xi_t
\end{equation}
\noindent
with $\alpha=0.5$ and the additive noise term $\xi_t$ taken as independent realizations of a Gaussian random variable with zero mean and unit variance, and (b) the $x$-component of the nonlinear-deterministic H\'enon map
\begin{equation}
x_t=A-x_{t-1}^2+By_{t-1}, \quad y_t=x_{t-1}
\end{equation}
\noindent
with $A=1.4$ and $B=0.3$. In both cases, we generate ensembles of independent realisations with random initial conditions and discard the first 1,000 points of each time series to avoid possible transients.

\begin{figure}[t]
\includegraphics[width=0.49\columnwidth]{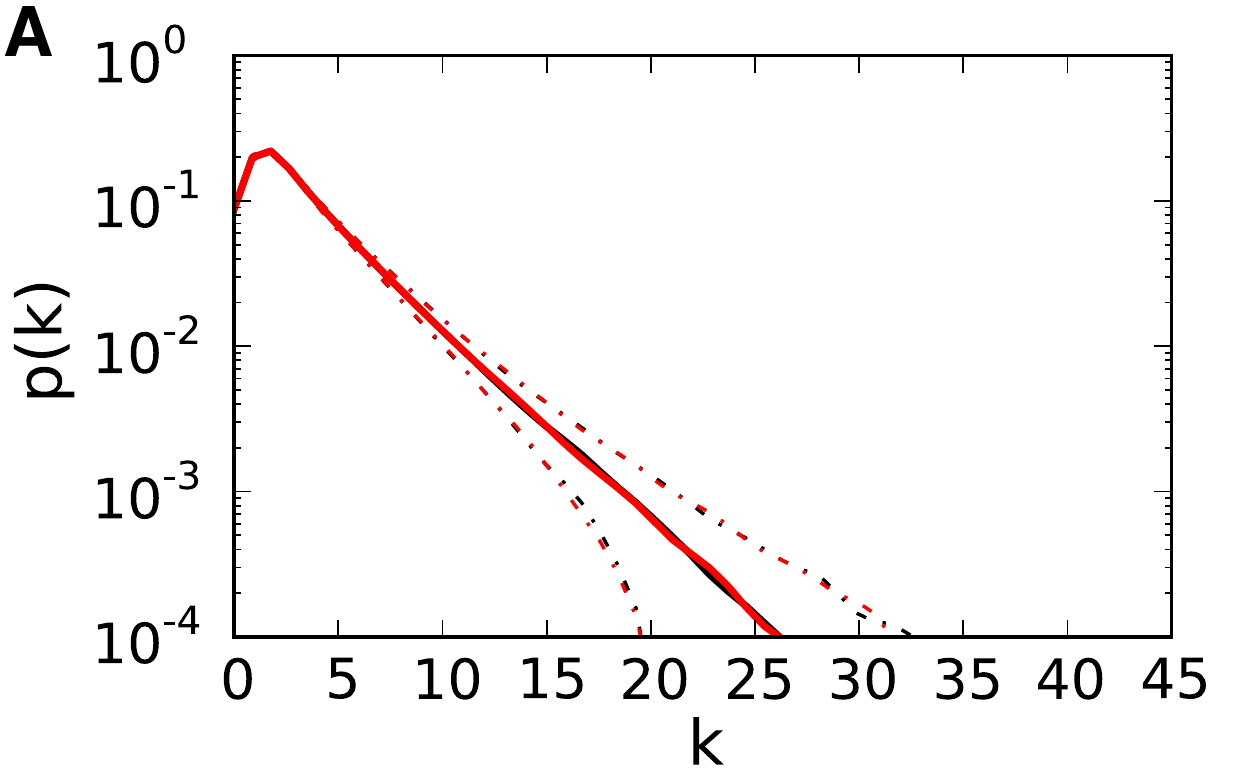} \hfill
\includegraphics[width=0.49\columnwidth]{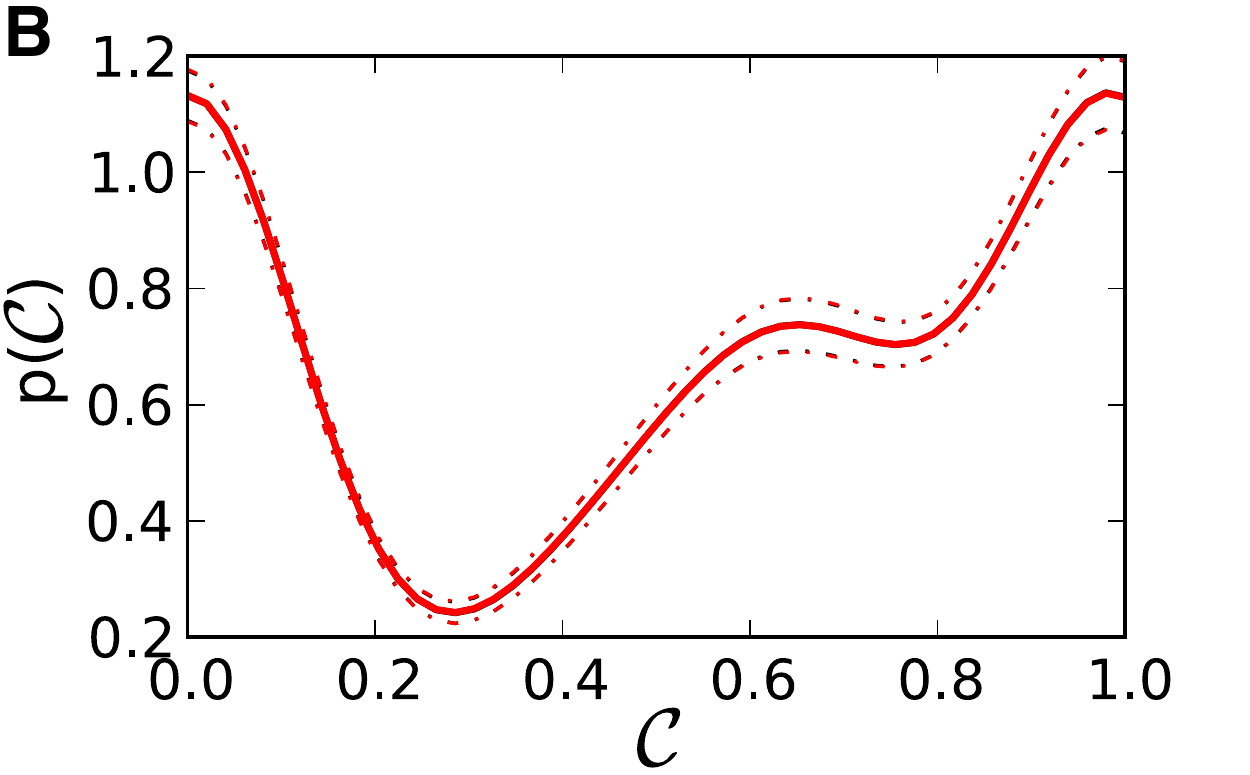} \\
\includegraphics[width=0.49\columnwidth]{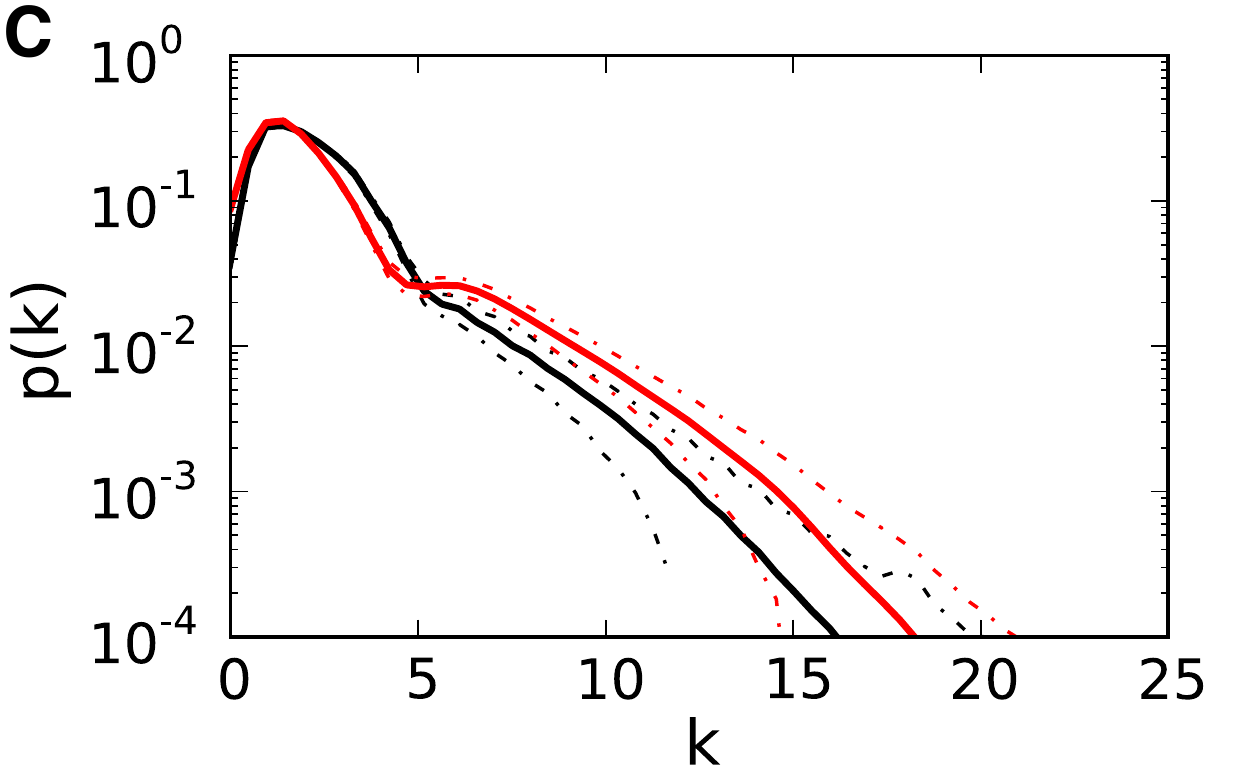} \hfill
\includegraphics[width=0.49\columnwidth]{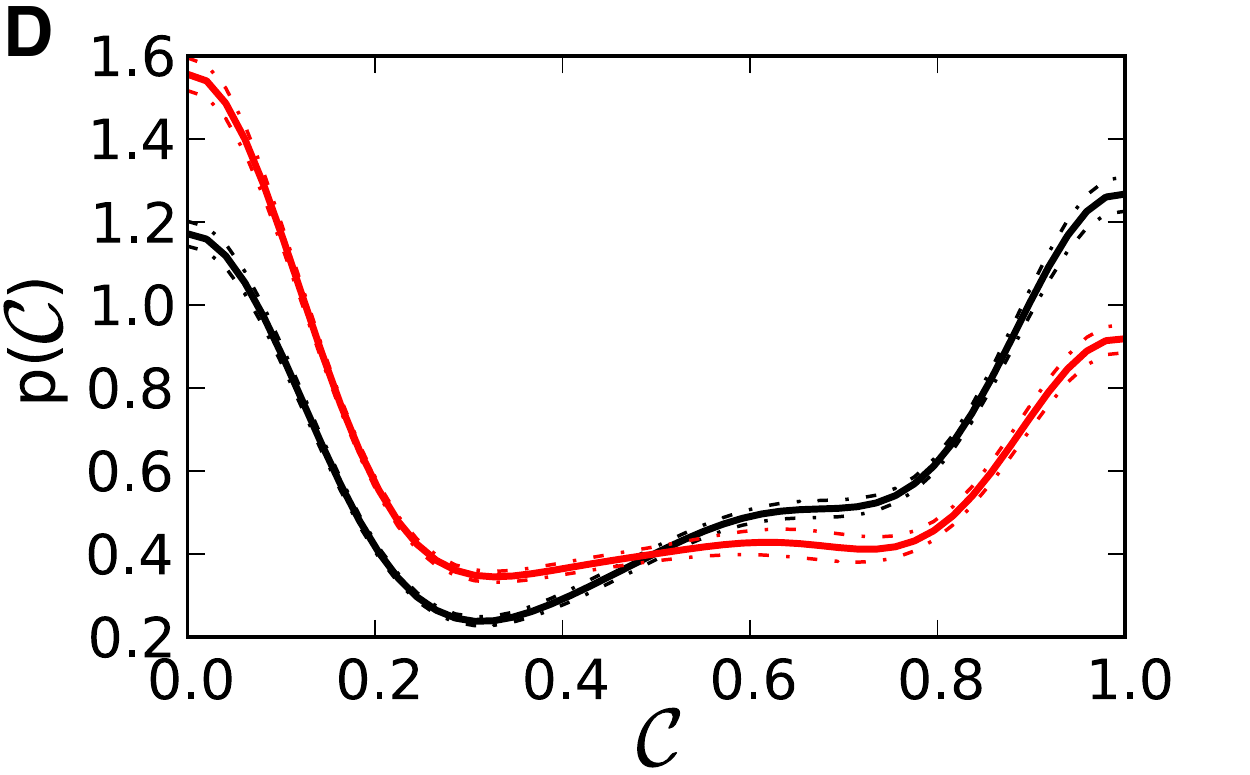} 
\caption{(Colour online) Distributions of retarded (black) and advanced (red) (A,C) degrees {$k_i^{r}$, $k_i^{a}$} and (B,D) local clustering coefficients {$\mathcal{C}^{r}_i$, $\mathcal{C}^{a}_i$} of the standard VG for two simple model systems: (A,B) AR(1) process and (C,D) H\'enon map ($x$-component). Time series of length $N=500$ have been used for estimating the probability density functions (PDF) {$p(k^{r})$, $p(k^{a})$, $p(\mathcal{C}^{r})$ and $p(\mathcal{C}^{a})$} with a kernel density estimator. The mean (solid lines) and standard deviation (dashed lines) of the PDFs have been computed based on an ensemble of $M=1,000$ realizations with random initial conditions for both systems.}
\label{examples}
\end{figure}

\begin{figure}[t]
\includegraphics[width=0.49\columnwidth]{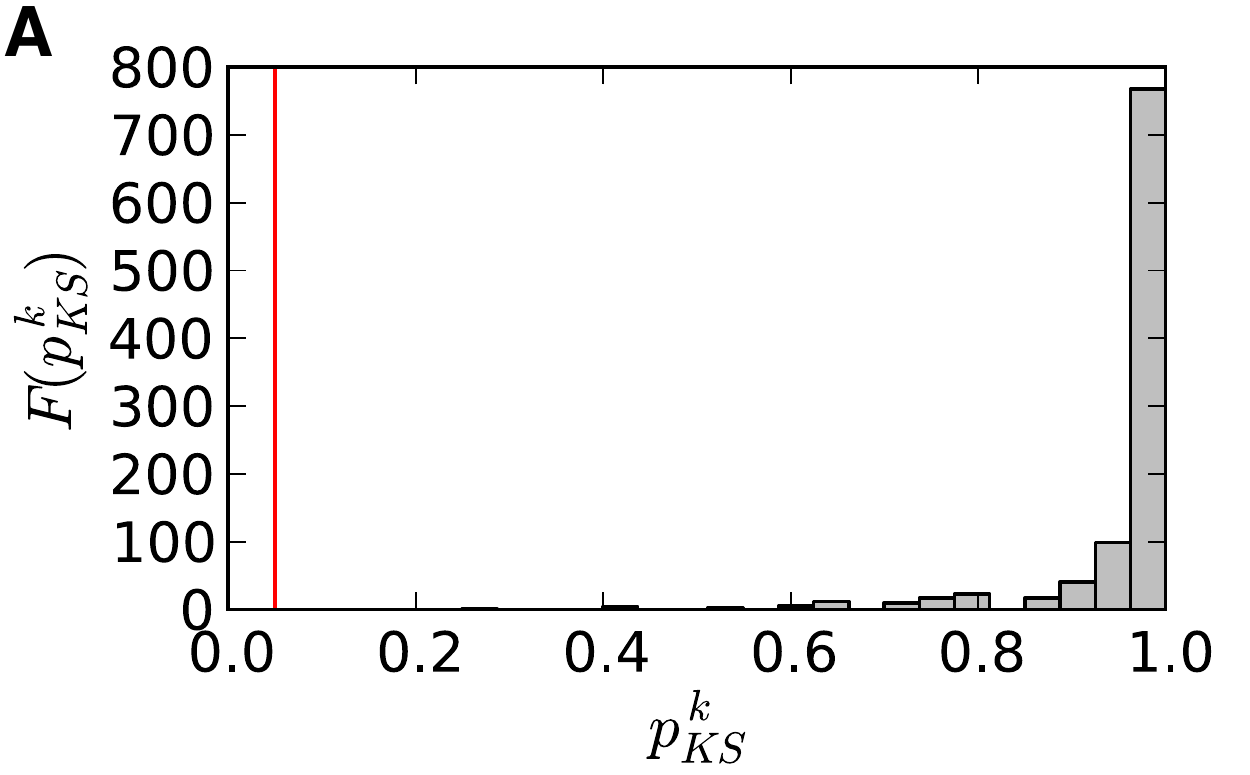} \hfill
\includegraphics[width=0.49\columnwidth]{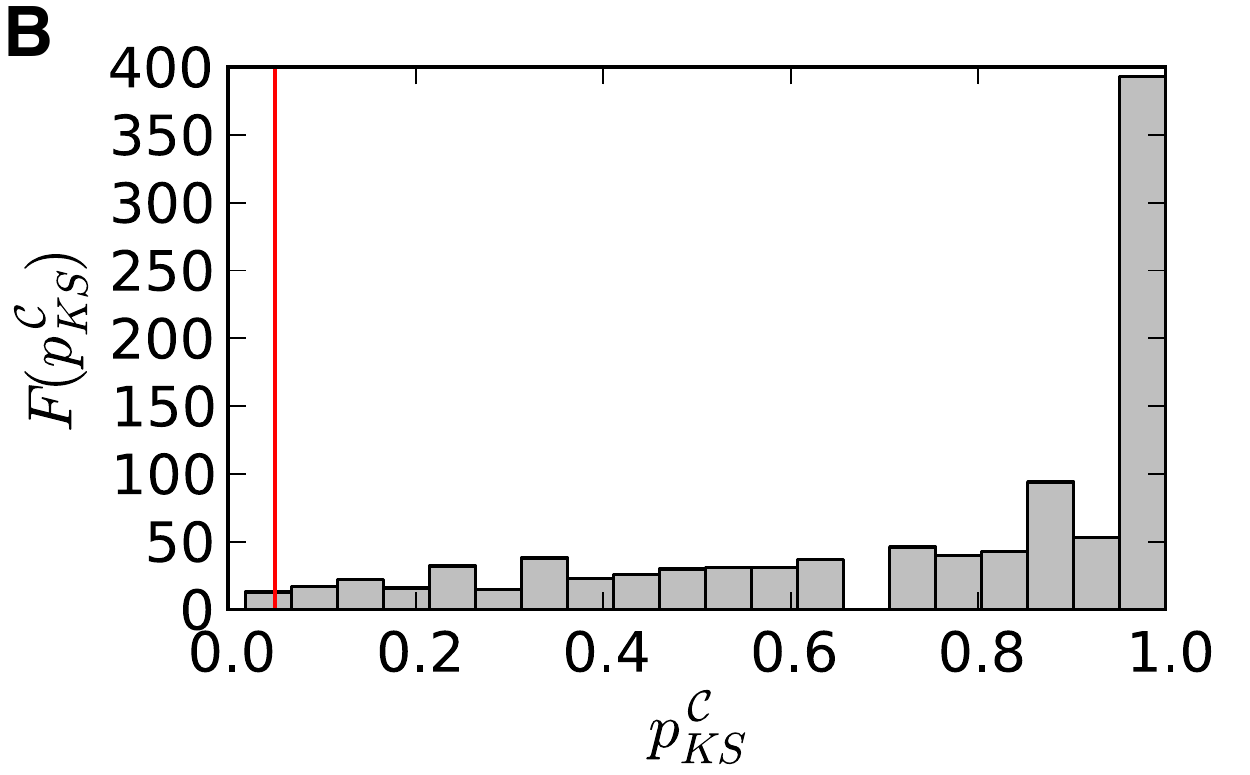} \\
\includegraphics[width=0.49\columnwidth]{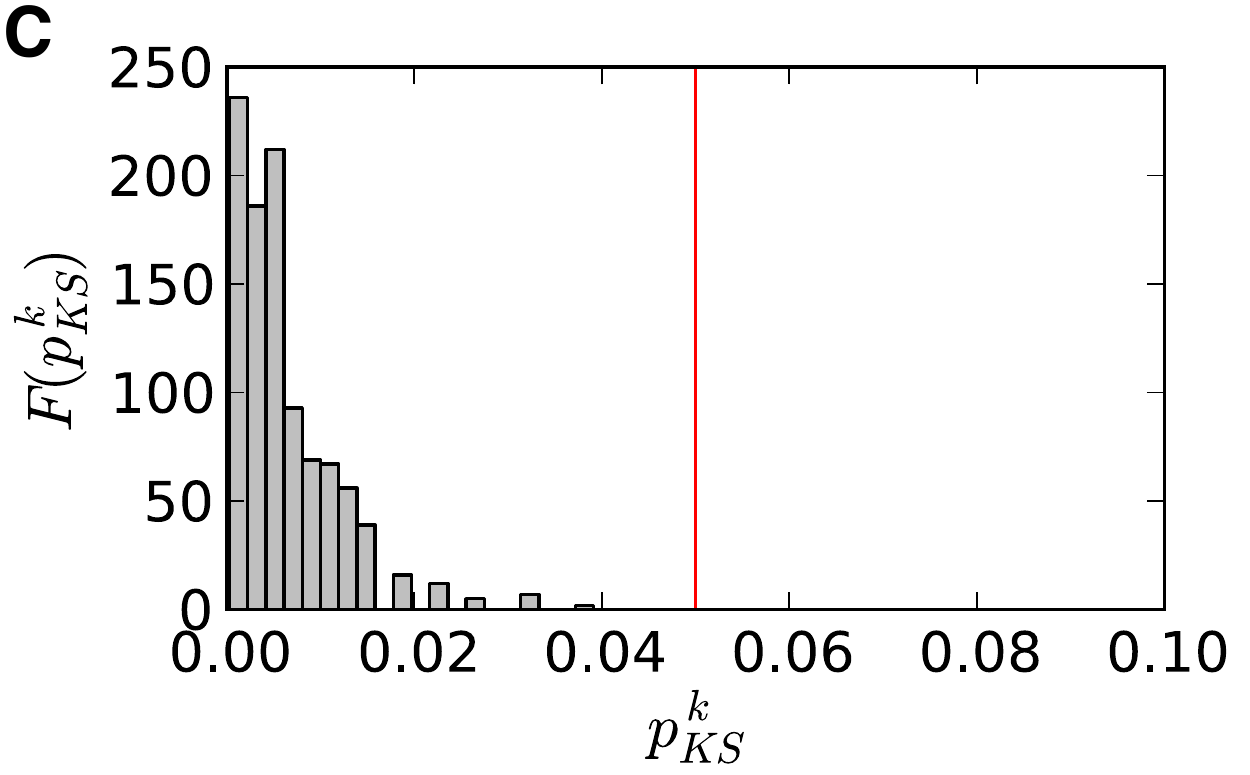} \hfill
\includegraphics[width=0.49\columnwidth]{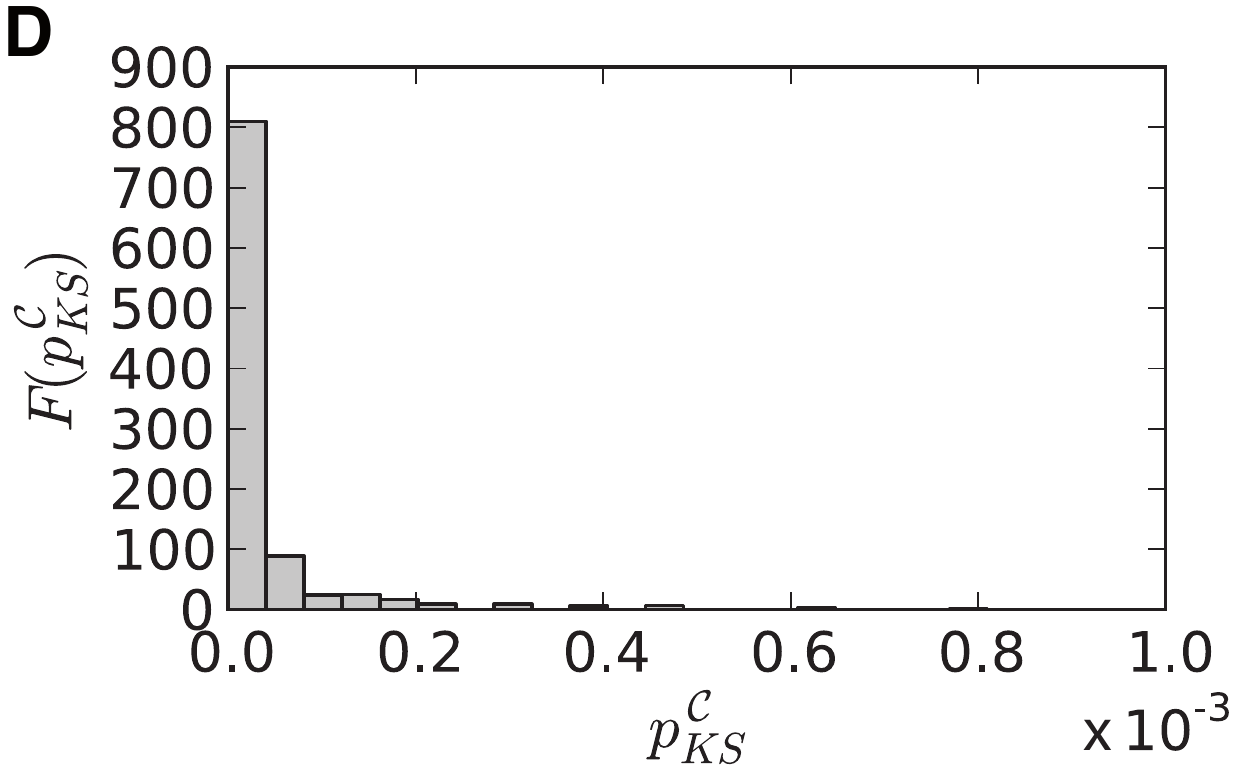}
\caption{(Colour online) Frequency distributions of $p$-values of the KS statistic for comparing the distributions of retarded/advanced (A,C) degree {$k_i^{r}$, $k_i^{a}$} and (B,D) local clustering coefficient {$\mathcal{C}^{r}_i$, $\mathcal{C}^{a}_i$} of standard VGs from an ensemble of $M=1,000$ realisations of model system time series of length $N=500$: (A,B) AR(1) process, (C,D) H\'enon map. Vertical red lines indicate the typical significance level of 0.05 {where appropriate (note the different scale in panel D)}.}
\label{examples2}
\end{figure}

As expected, for the linear (reversible) AR(1) process, the empirical distributions of retarded/advanced {vertex properties} collapse onto each other (Fig.~\ref{examples}A,B). Consequently, the null hypothesis of reversibility is never rejected by the test based on the degree (Fig.~\ref{examples2}A), and only rarely rejected by the clustering-based test well below the expected false rejection rate of 5\% (Fig.~\ref{examples2}B). Similar results are obtained for AR1 realizations very close to Brownian motion with $\alpha=0.9$ {and $0.99$}. In contrast, for the irreversible H\'enon map the distributions of retarded and advanced VG measures appear distinct already by visual inspection (Fig.~\ref{examples}C,D). In accordance with this observation, the null hypothesis of reversibility is nearly always (degree, Fig.~\ref{examples2}C) or always (local clustering coefficient, Fig.~\ref{examples2}D) rejected. Consistently, an even higher rejection rate is found for the highly nonlinear and infinite dimensional Mackey-Glass system~\cite{Mackey1977} in periodic and hyperchaotic regimes. All results are qualitatively independent of the chosen network construction algorithm (VG or HVG).

\begin{figure}
\includegraphics[width=0.49\columnwidth]{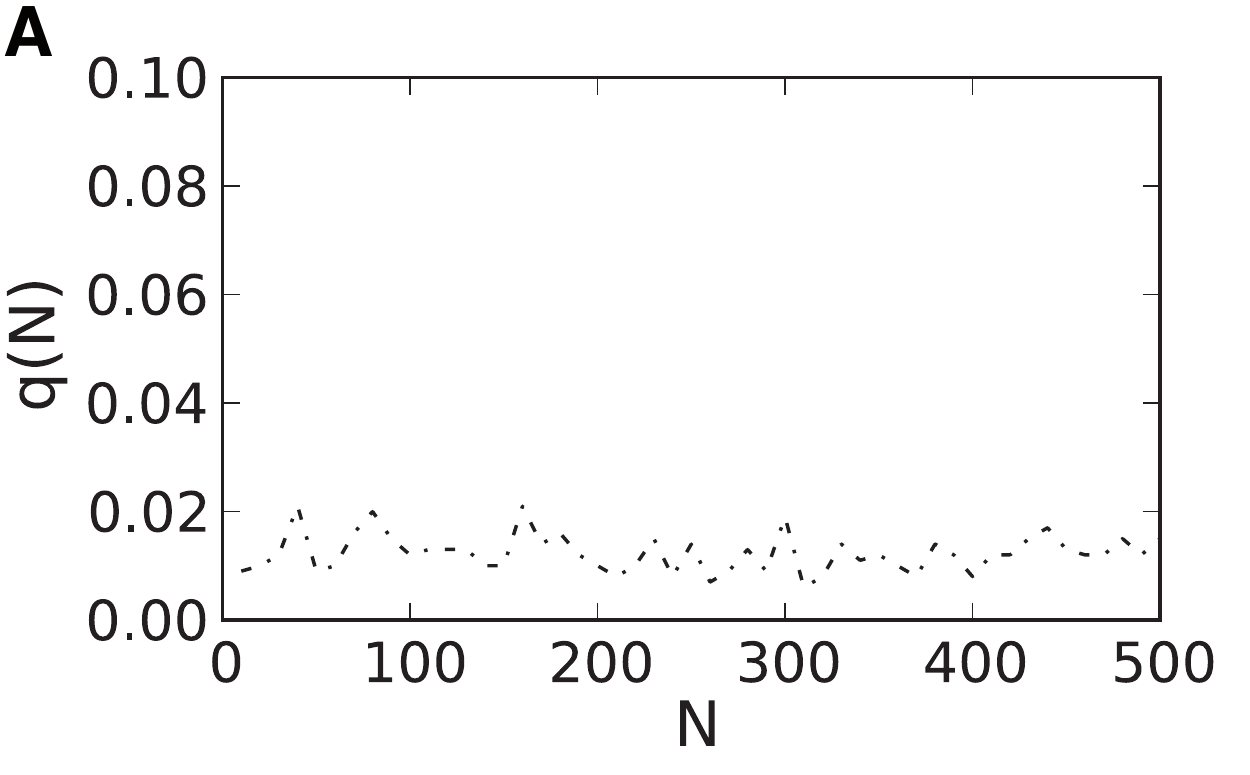} \hfill
\includegraphics[width=0.49\columnwidth]{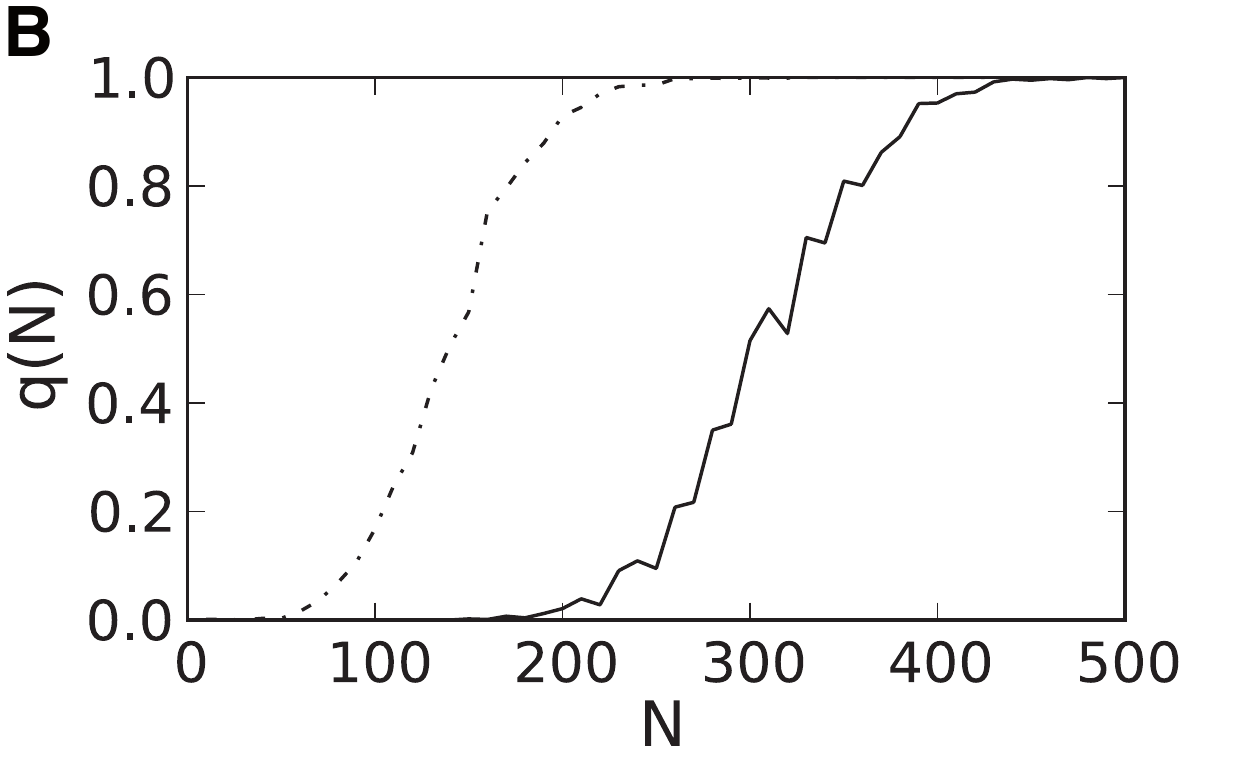} \\ \includegraphics[width=0.49\columnwidth]{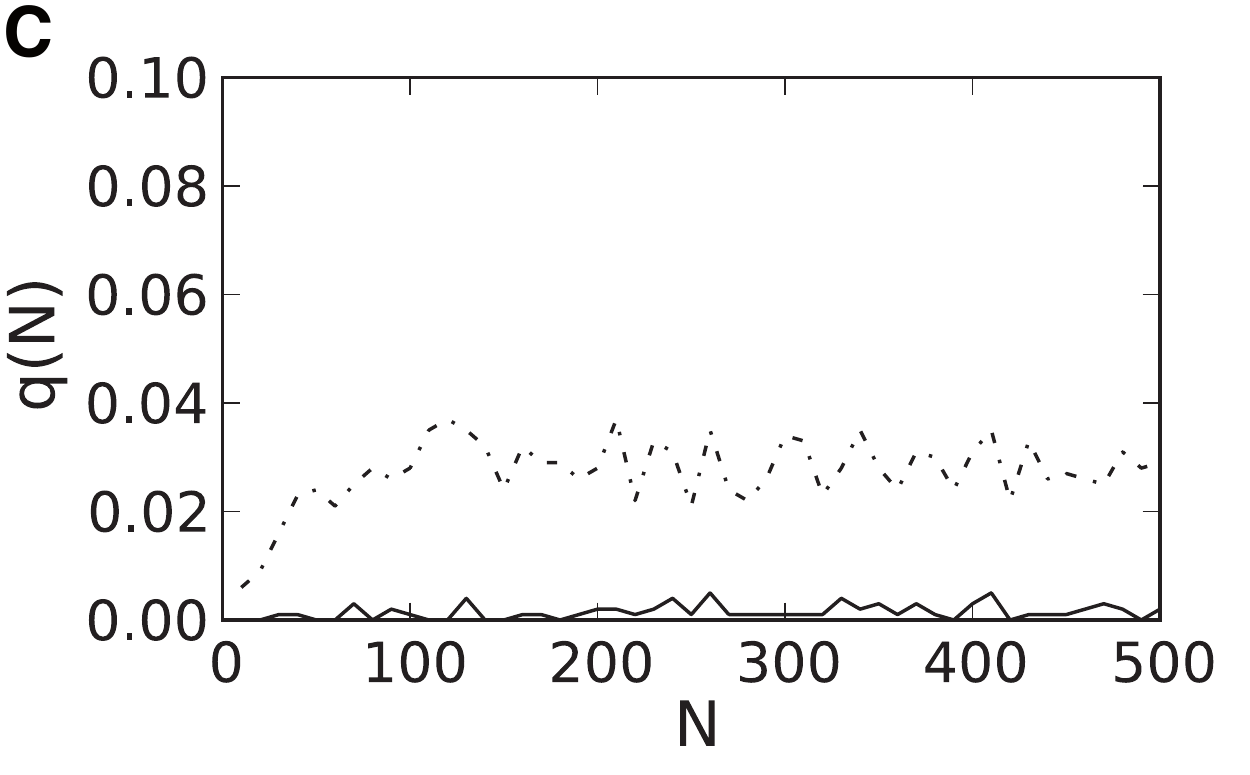} \hfill
\includegraphics[width=0.49\columnwidth]{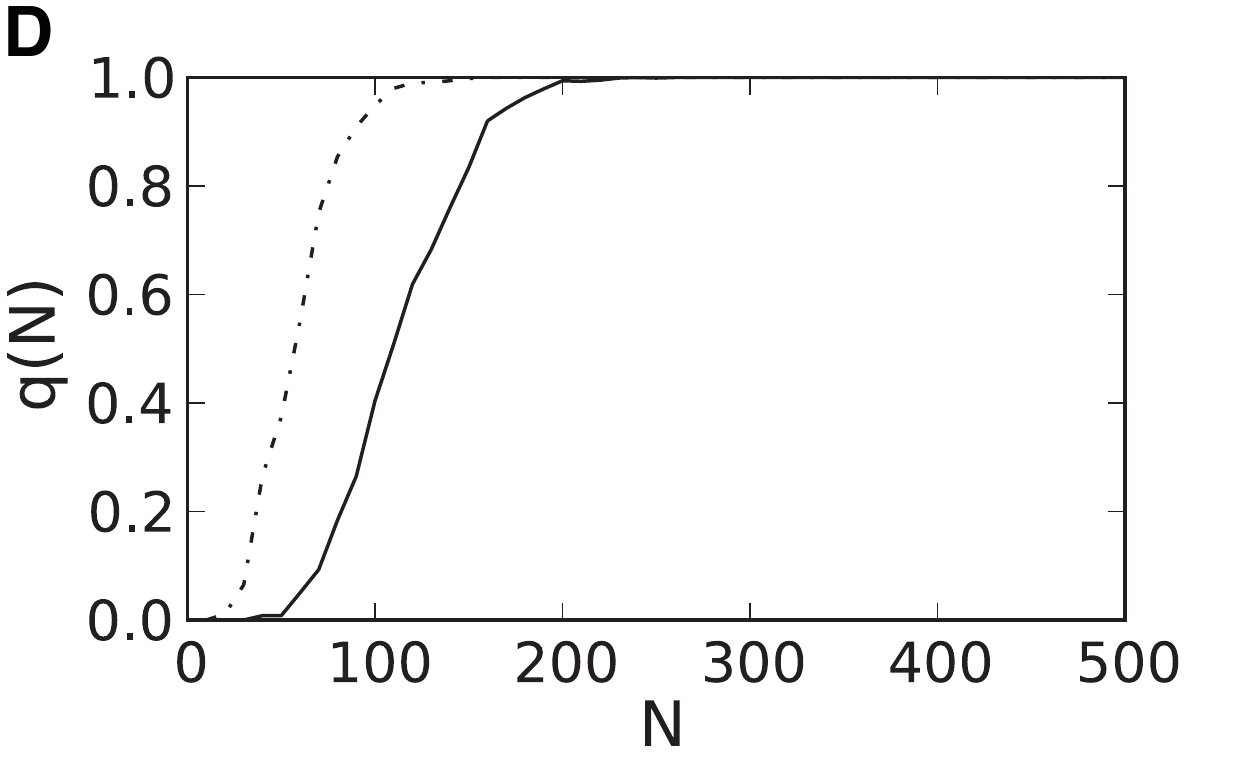}
\caption{Fraction $q(N)$ of model time series of length $N$ from ensembles of $M=1,000$ realisations for which the null hypothesis of reversibility was rejected at the 0.05 significance level: (A,C) AR(1) process, (B,D) H\'enon map. The results have been obtained using (A,B) VGs and (C,D) HVGs {with degree-} (solid lines) and {clustering-based tests} (dash-dotted lines), respectively. The null hypothesis is never rejected for the {VG degree-}based test applied to the AR(1) time series (A).}
\label{examples3}
\end{figure}

To further evaluate the performance of the tests for varying sample size $N$, we consider the fraction $q(N)$ of time series from an ensemble of realisations for which the null hypothesis of reversibility can be rejected (Fig.~\ref{examples3}). For the AR(1) process, it is known that the null hypothesis is true. Hence, $q(N)$ estimates the probability of \emph{type I errors} (incorrect rejections of true null hypothesis) for both tests (Fig.~\ref{examples3}A,C). To put it differently, $1-q(N)$ measures the \emph{specificity} of the test. Notably, for the standard VG, $q(N)$ is always zero for the degree-based test, while it fluctuates clearly below the expected type I error rate of 0.05 for the {clustering-based} test (Fig.~\ref{examples3}A). For the HVG-based tests, $q(N)$ takes slightly higher values, which, however, remain below the acceptable error level (Fig.~\ref{examples3}C).

In contrast to the linear AR(1) process, for the irreversible H\'enon map the null hypothesis of reversibility is known to be false. Therefore, $q(N)$ measures the power of the test, whereas $1-q(N)$ gives the probability of \emph{type II errors} (failure to reject a false null hypothesis), i.e., its \emph{sensitivity} (Fig.~\ref{examples3}B,D). Interestingly, the power of the clustering-based test increases markedly earlier than that of the degree-based test. For the standard VG-based test (Fig.~\ref{examples3}B), the former reaches $q(N)\approx 1$ already around $N=200$, whereas the latter requires twice as many samples to arrive at the same power. Notably, the convergence towards $q(N)=1$ is much faster for the HVG algorithm (Fig.~\ref{examples3}D), leading to a perfect hit rate at $N=100$ and $200$ for local clustering coefficients and degrees, respectively.

\begin{figure}
\includegraphics[width=0.49\columnwidth]{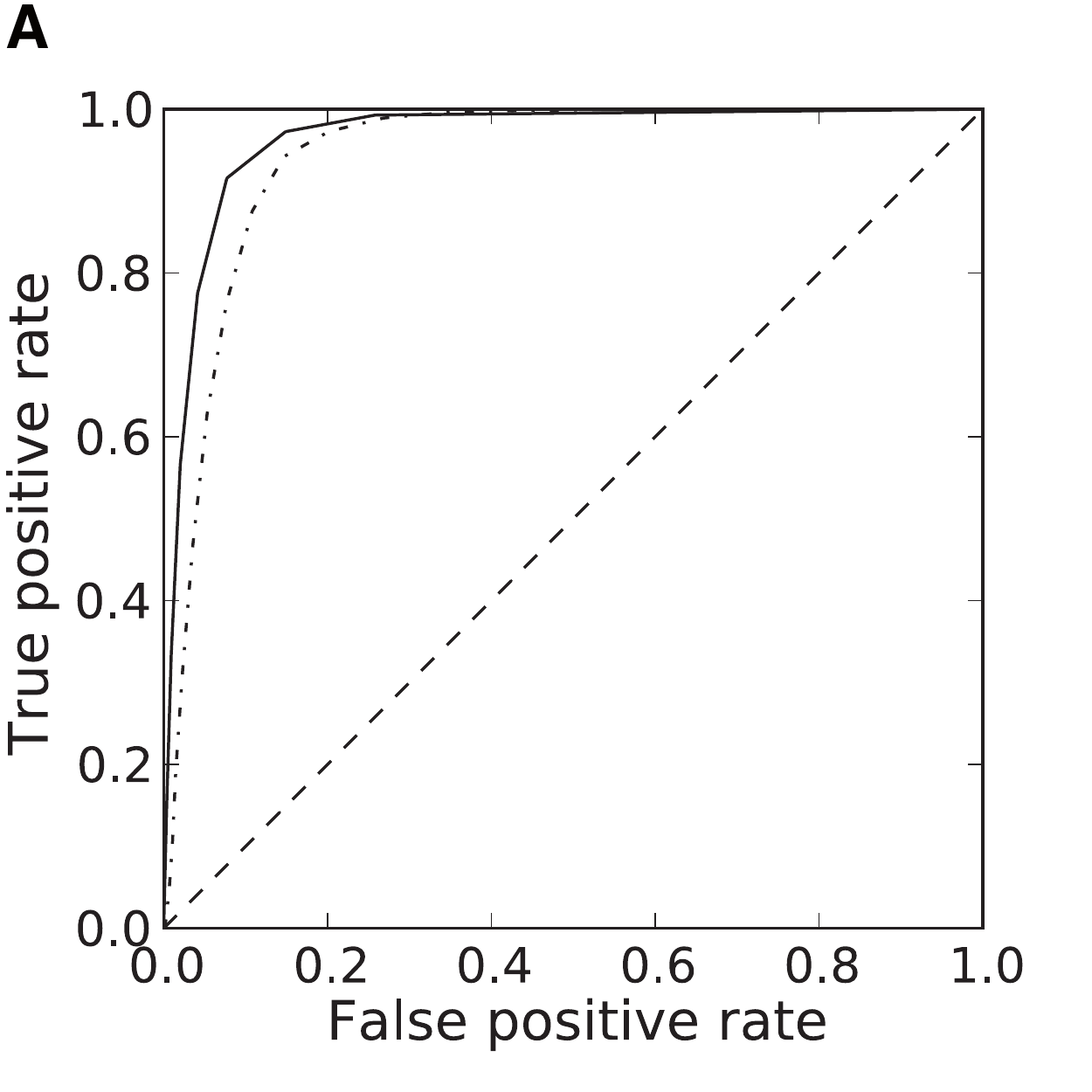} \hfill
\includegraphics[width=0.49\columnwidth]{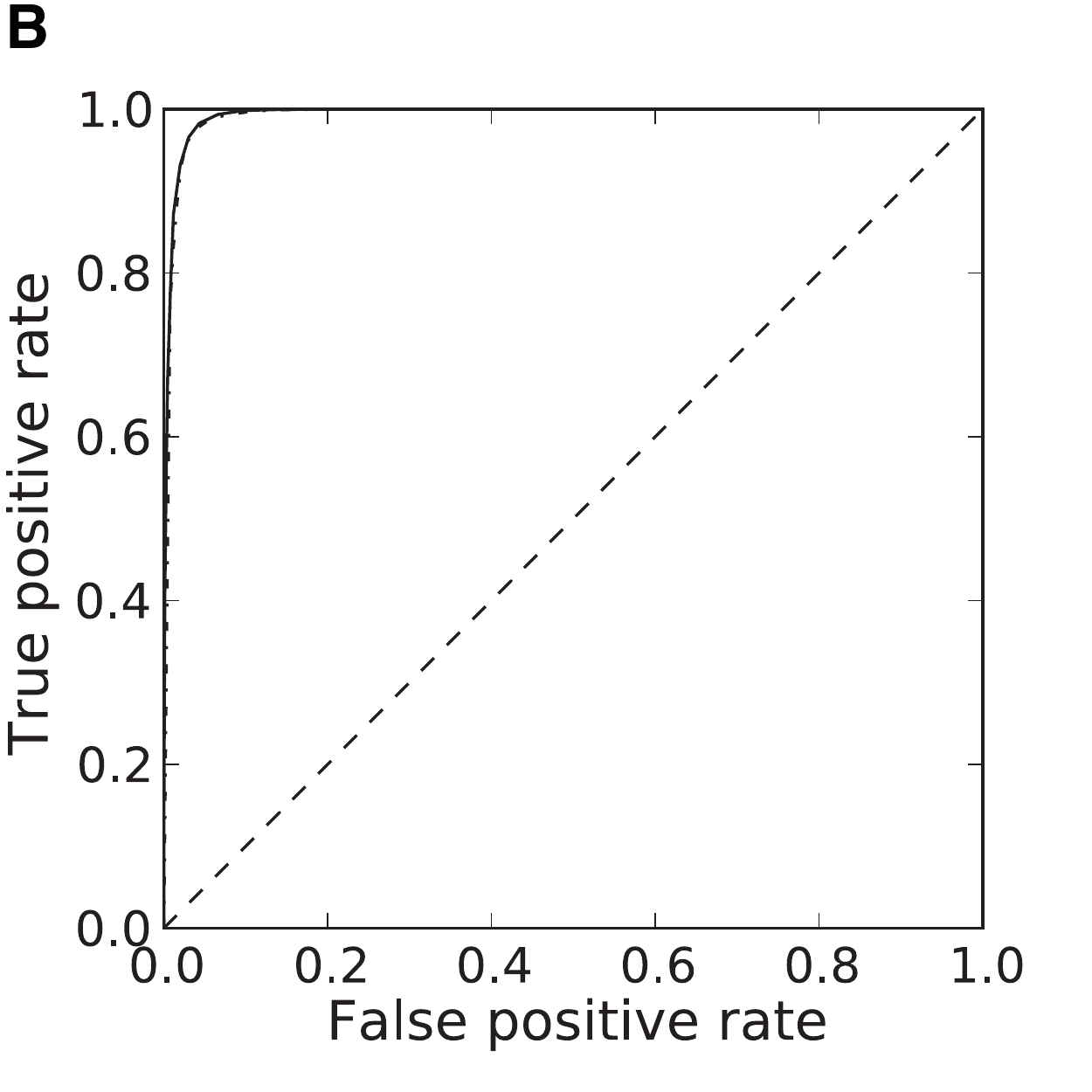} \\
\includegraphics[width=0.49\columnwidth]{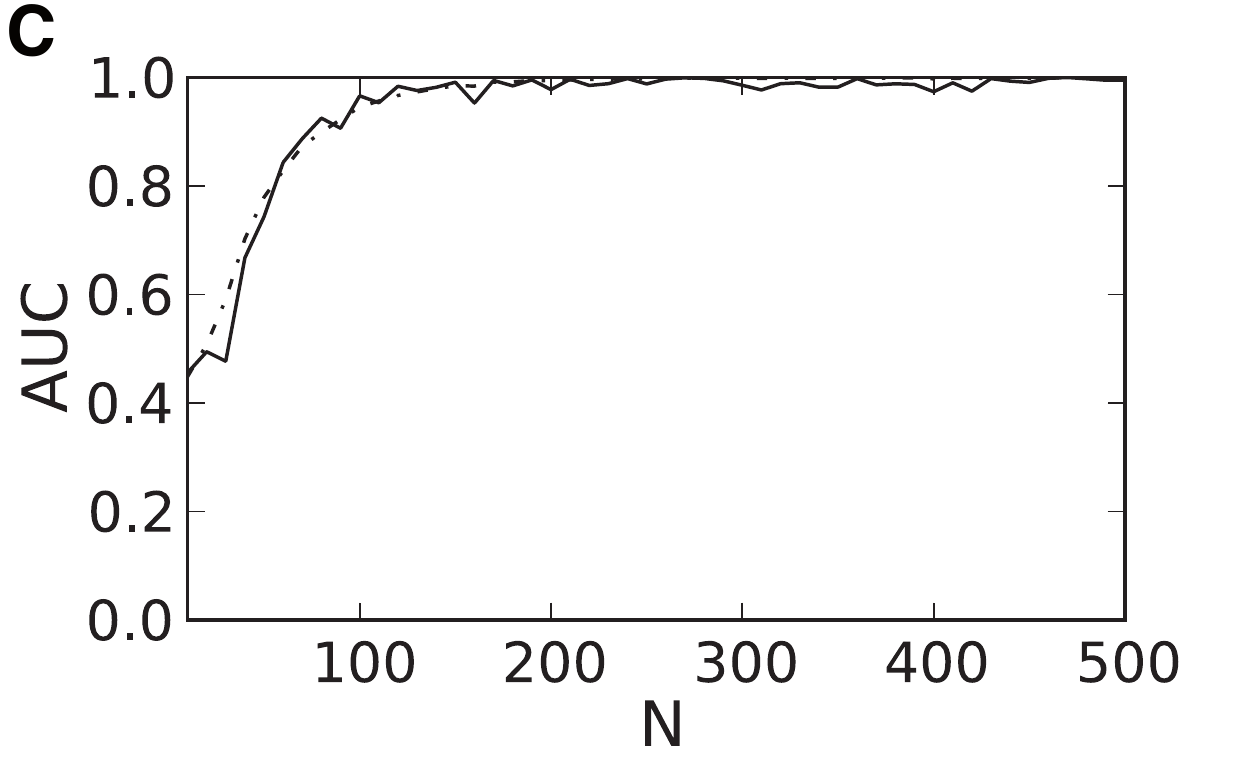} \hfill
\includegraphics[width=0.49\columnwidth]{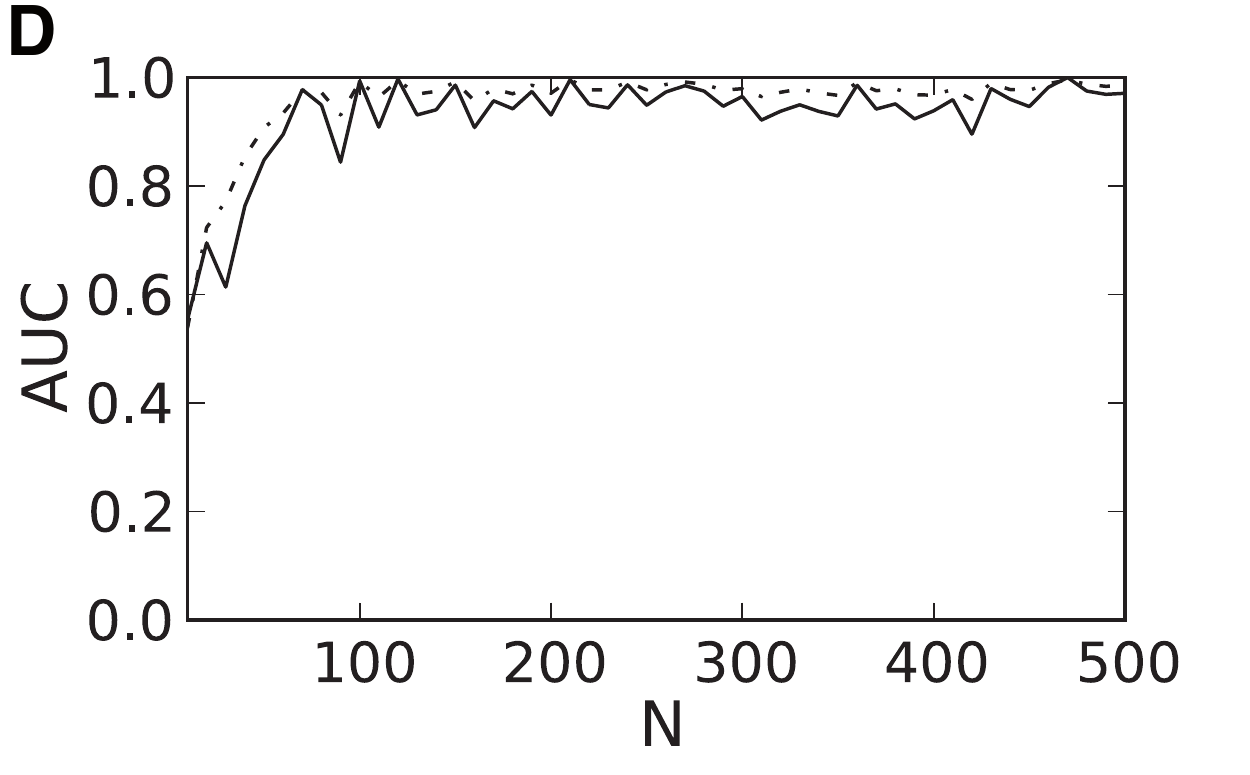}
\caption{(A,B) ROC curves for the (A,C) VG- and (B,D) HVG-based tests for reversibility comparing the rejection rates for each $M=10,000$ realisations of AR1 (false positive rate) and H\'enon time series (true positive rate) with varying critical $p$-value of the KS statistic ($N=100$). (C,D) Area under the ROC curve (${\text{AUC}}$) characterising the discriminative performance of all tests depending on time series length $N$. Solid and dash-dotted lines indicate degree- and {clustering-based} tests, respectively.}
\label{examples4}
\end{figure}

Receiver operating characteristics (ROC curves) enable a more systematic comparison between the VG- and HVG-based tests. For varying the critical $p$-value of the KS statistic for rejecting the reversibility hypothesis, the rejection probability for AR1 (false positive rate) and H\'enon (true positive rate) time series is plotted. Given a fixed true positive rate, the VG-based tests typically have a higher error probability (false positive rate) than those utilising HVGs (Fig.~\ref{examples4}A,B). As the length $N$ of the individual records increases, VG-based tests display better convergence properties towards the ideal ROC curve with area under ROC curve ${\text{AUC}}=1$, whereas the HVG-based tests show substantial fluctuations even for relatively long time series (Fig.~\ref{examples4}C,D). We attribute the faster convergence, but larger residual error probability of the HVG-based tests to the stronger constraints imposed during network construction in comparison with the standard VG algorithm. Since the VG involves more edges than the HVG, the former is more robust and less resilient to statistical fluctuations, but a larger number of vertices is necessary for identifying irreversible behaviour in the data under study. Notably, particularly for HVGs, the {clustering-based} tests systematically perform slightly better than those using degree (Fig.~\ref{examples4}D).

\section{Real-world example}

To further demonstrate the potentials of {(H)}VG-based {ir}reversibility tests for real-world data, we apply them to continuous electroencephalogram (EEG) recordings for healthy and epileptic patients that were previously analysed by Andrzejak~\textit{et~al.}~\cite{Andrzejak2001}. The data consist of five sets of $M=100$ representative time series segments of length $N=4,096$ comprising recordings of brain activity for different patient groups and recording regions (Table~\ref{tab:eeg_vg_reversibility_test}). To look for traces of low-dimensional nonlinear dynamical behaviour in the data, Andrzejak~\textit{et~al.}~\cite{Andrzejak2001} used the nonlinear prediction error $P$ and the effective correlation dimension $D_{2,{\text{eff}}}$ as statistics to test the null hypothesis $H_0^{{\text{lin}}}$ that the time series are compatible with a stationary linear-stochastic Gaussian process.

\begin{table*}[tb]
\caption{Results of {(H)}VG-based tests for {ir}reversibility of EEG time series put into context with the results from~\cite{Andrzejak2001} based on the nonlinear prediction error $P$ and the effective correlation dimension $D_{2,{\text{eff}}}$ as statistics to test the null hypothesis $H_0^{{\text{lin}}}$ that the time series are compatible with a stationary linear-stochastic Gaussian process. $q_k$ ($q_\mathcal{C}$) denotes the fraction of time series from a set of $M=100$ segments for which the null hypothesis of reversibility was rejected by the proposed {(H)}VG-based tests using the retarded/advanced degrees (local clustering coefficients) (see also Fig.~\ref{fig:p_vals_eeg}).}
\vspace{0.4cm}
\centering
\begin{tabular}{lllllrrrr}
\hline
& & & \multicolumn{2}{l}{Andrzejak~\textit{et~al.}~\cite{Andrzejak2001}} & \multicolumn{2}{l}{VG} & \multicolumn{2}{l}{HVG} \\
\hline
Set & State & Recording sites & $P$ & $D_{2,{\text{eff}}}$ & $q_k$ & $q_\mathcal{C}$ & $q_k$ & $q_\mathcal{C}$ \\
\hline
A & healthy,  & mixed & no reject $H_0^{{\text{lin}}}$ & no reject $H_0^{{\text{lin}}}$ & $0.00$ & $0.07$ & $0.01$ & $0.08$ \\
& eyes open \\
B & healthy, & mixed & reject $H_0^{{\text{lin}}}$ & no reject $H_0^{{\text{lin}}}$ & $0.07$ & $0.16$ & $0.17$ & $0.37$ \\
 & eyes closed \\
C & pathological, & hippocampal & reject $H_0^{{\text{lin}}}$ & no reject $H_0^{{\text{lin}}}$ & $0.13$ & $0.22$ & $0.10$ & $0.21$ \\
 & no seizure & formation \\
D & pathological, & epileptogenic zone & reject $H_0^{{\text{lin}}}$ & reject $H_0^{{\text{lin}}}$ & $0.36$ & $0.37$ & $0.35$ & $0.55$ \\
 & no seizure \\
E & pathological, & mixed & reject $H_0^{{\text{lin}}}$ & reject $H_0^{{\text{lin}}}$ & $0.87$ & $0.94$ & $0.87$ & $0.93$ \\
 & seizure \\
\hline
\end{tabular}
\label{tab:eeg_vg_reversibility_test}
\end{table*}%

\begin{figure}[tb]
   \centering
   \includegraphics[width=0.49\columnwidth]{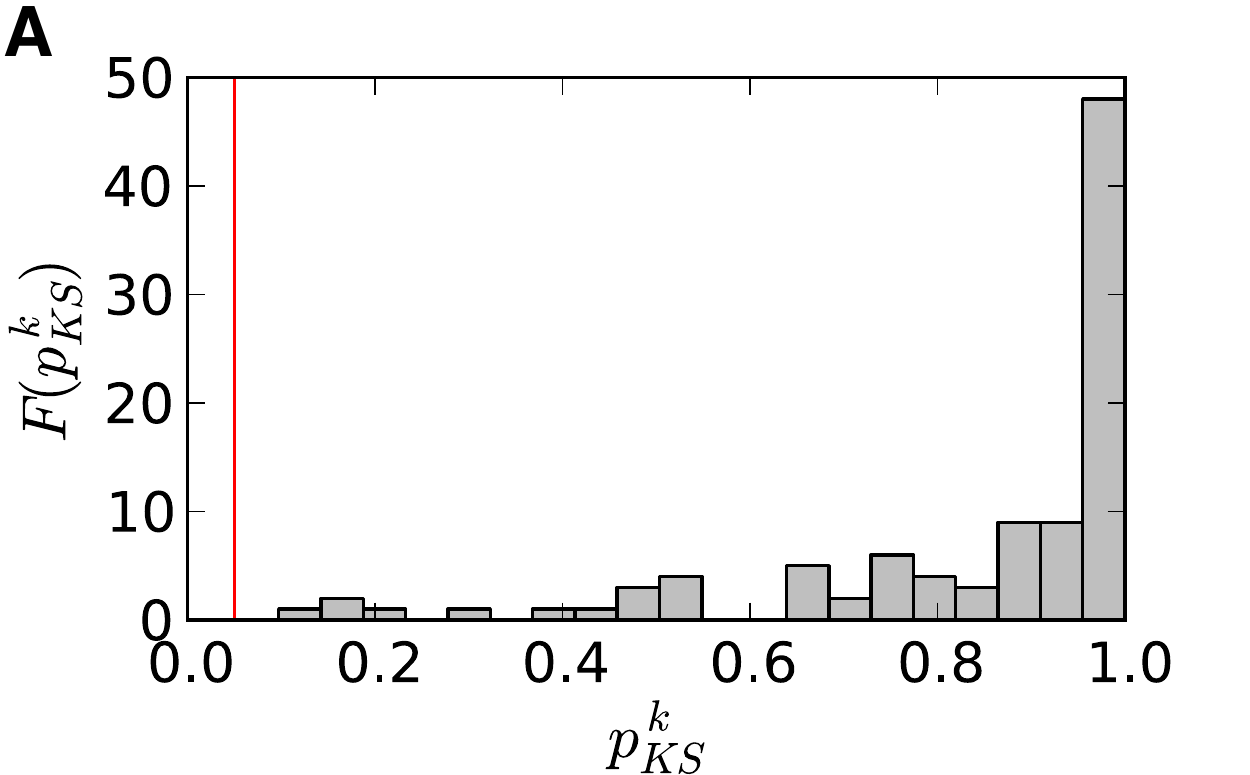}
   \includegraphics[width=0.49\columnwidth]{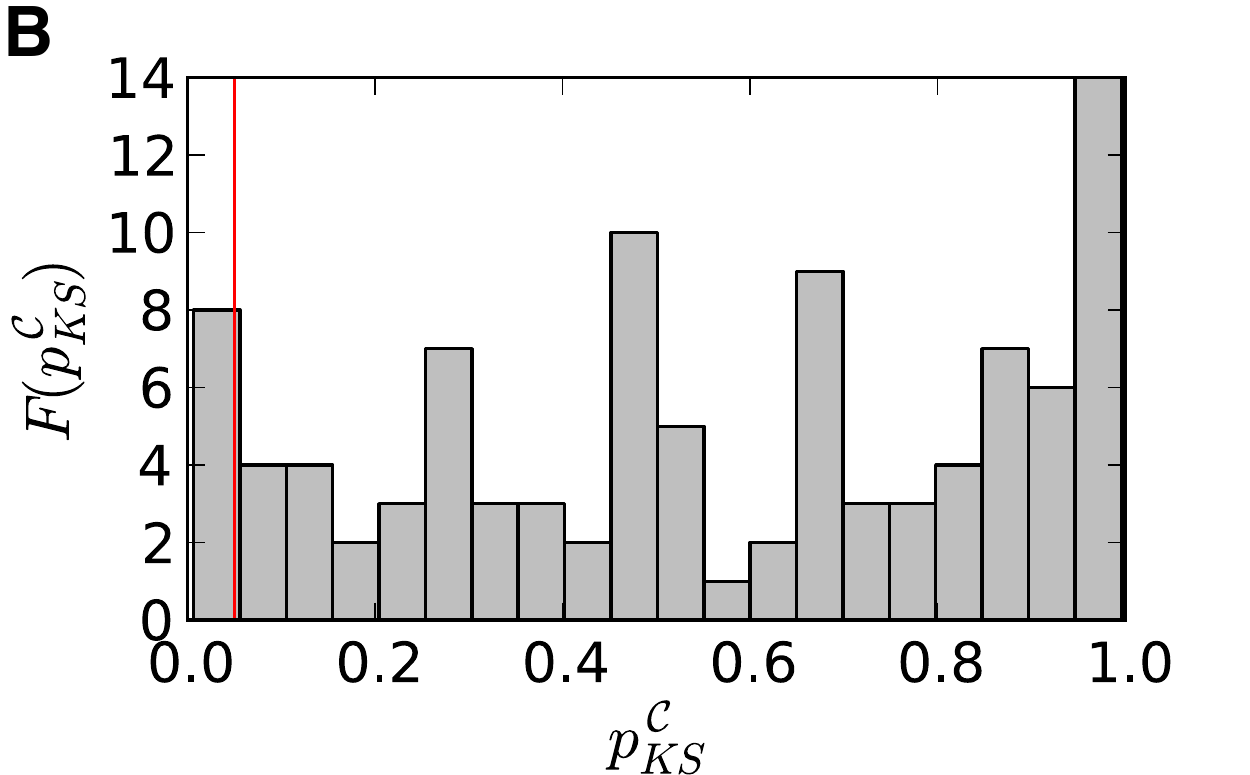}
   \includegraphics[width=0.49\columnwidth]{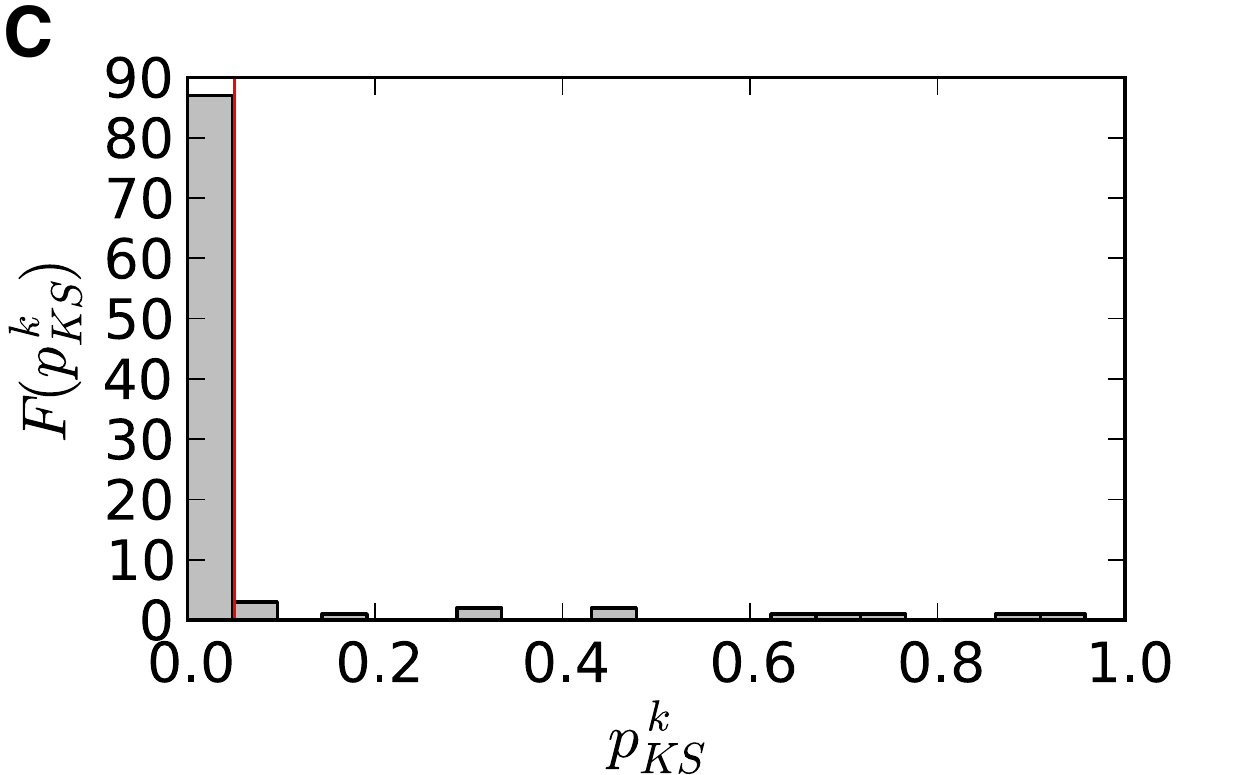}
   \includegraphics[width=0.49\columnwidth]{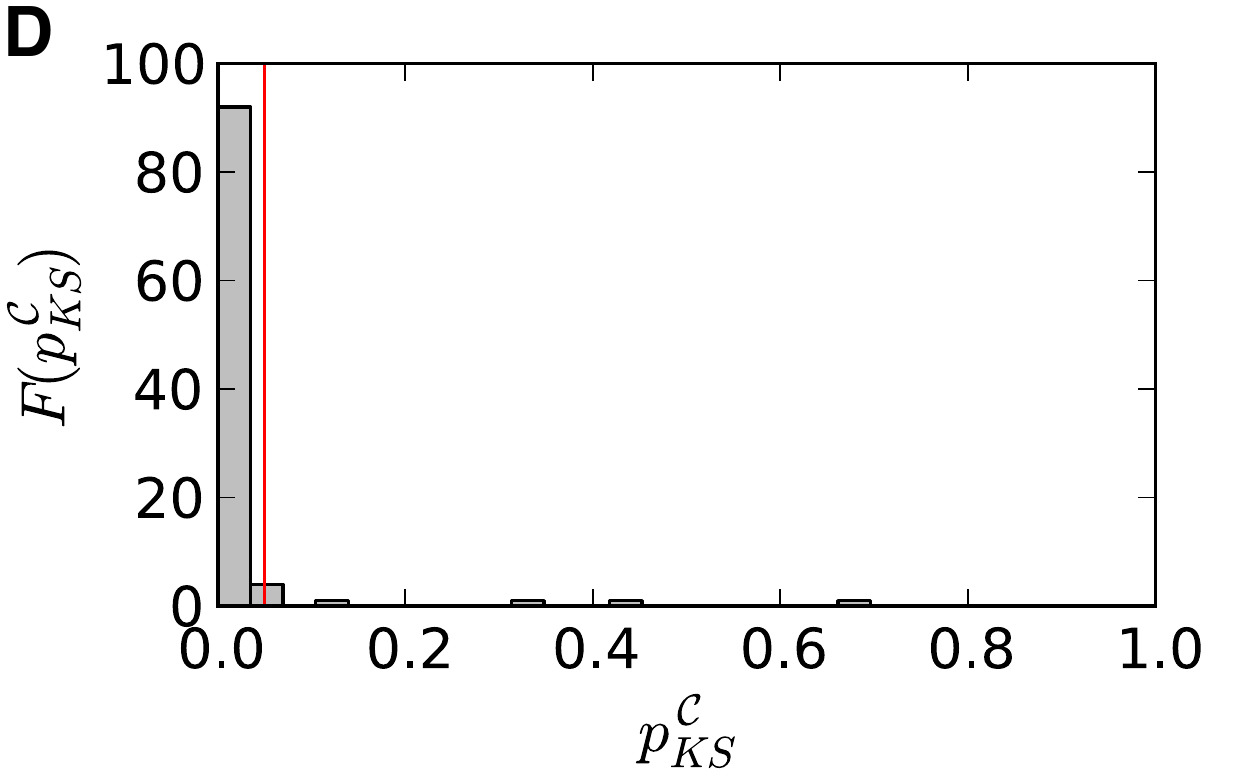}
   \caption{Frequency distributions of $p$-values of the KS test for comparing the distributions of retarded/advanced (A,C) degree {$k_i^{r}$, $k_i^{a}$} and (B,D) local clustering coefficient {$\mathcal{C}_i^{r}$, $\mathcal{C}_i^{a}$} of standard VGs from a set of $M=100$ EEG time series segments of length $N=4,096$. Recordings originate from (A,B) healthy subjects with eyes open (data set A) and (C,D) epileptic patients during seizure (data set E). Vertical red lines indicate the chosen significance level of 0.05.}
   \label{fig:p_vals_eeg}
\end{figure}

Since irreversibility is a signature of nonlinear dynamics, we expect the results of our {(H)}VG-based tests to be consistent with those of~\cite{Andrzejak2001}. Indeed, the rate of rejections $q$ of the null hypothesis of reversibility increases markedly from hardly any rejections for set A, where $H_0^{{\text{lin}}}$ could not be rejected by~\cite{Andrzejak2001}, to $q\approx 1$ for set E, where $H_0^{{\text{lin}}}$ was rejected using both test statistics (Table~\ref{tab:eeg_vg_reversibility_test}). Hence, consistently with the results of~\cite{Andrzejak2001}, the {(H)}VG-based tests indicate {probably} reversible dynamics for healthy subjects (set A, Fig.~\ref{fig:p_vals_eeg}A,B) and {clearly} irreversible {(}nonlinear{)} dynamics during epileptic seizures (set E, Fig.~\ref{fig:p_vals_eeg}C,D). The other data sets (B-D) are identified as intermediate cases with respect to the proposed tests, suggesting time-reversal asymmetry and, hence, nonlinear dynamics.

In summary, our tests perform consistently with those applied by~\cite{Andrzejak2001} which are arguably more complicated both technically and conceptually. Furthermore, the results of the {(H)}VG-based tests are consistent with those obtained using {a third-order} statistics~\cite[p. 84]{Theiler1992} together with standard and amplitude adjusted Fourier {surrogates}~\cite{Schreiber2000}, a {classical} test for time series {ir}reversibility.

\section{Conclusions}

The statistical tests for {ir}reversibility of scalar-valued time series proposed in this work {provide an example for the wide applicability} of complex network-based approaches for time series analysis problems. Utilising standard as well as horizontal VGs for discriminating between the properties of observed data forwards and backwards in time has at least two important {benefits}: 

(i) {Unlike for some classical tests (e.g.,~\cite{Theiler1992}),} the reversibility properties are examined without the necessity of constructing surrogate data. Hence, the proposed approach saves considerable computational costs in comparison with {such methods} and, more importantly, avoids the problem of selecting a particular type of surrogates. Specifically, utilising the KS test statistic or a comparable two-sample test for the homogeneity (equality) of the underlying probability distribution functions directly supplies a $p$-value for the associated null hypothesis that the considered properties of the data forward and backward in time are statistically indistinguishable. 

(ii) The proposed approach {is applicable} to data with non-uniform sampling {(common in areas like} palaeoclimate~\cite{Donner2012AG} or astrophysics) {and} marked point processes (e.g., earthquake catalogues~\cite{Telesca2012}). For {such} data, constructing surrogates for {non}linearity tests in the most common way {using} Fourier-based techniques is a challenging task, {which is avoided by} {(H)}VG-based methods. 

{We emphasise that our method exploits the time-information explicitly used in constructing (H)VGs. Other existing time series network methods (e.g., recurrence networks~\cite{Marwan2009,Donner2010NJP,Donner2011IJBC}) not exhibiting this feature cannot be used for the same purpose.}

While this {Letter} highlights the potentials of the proposed approach, there are methodological questions such as the impacts of sampling, observational noise, {and intrinsic correlations in vertex characteristics} as well as a systematic comparison to existing methods for testing time series {ir}reversibility that need to be systematically addressed in future research. Furthermore, {(H)}VG-based methods are generally faced with problems such as boundary effects and the ambiguous treatment of missing data~\cite{Donner2012AG}, which call for further investigations. 

Finally, {we note that} other measures characterising complex networks on the local (vertex/edge) as well as global scale could be used for similar purposes as {those} studied in this work. However, {since} path-based network characteristics {(e.g.,} closeness, betweenness, or average path length{)} cannot be easily decomposed into retarded and advanced contributions, the approach followed here is mainly restricted to neighbourhood-based network measures like degree, local and global clustering coefficient, or network transitivity. As a possible solution, instead of decomposing the network properties, the whole edge set of a (H)VG could be divided into two disjoint subsets that correspond to visibility connections forwards and backwards in time, as originally proposed by Lacasa~\textit{et~al.}~\cite{Lacasa2012}. For these directed (forward and backward) (H)VGs, also the path-based measures can be computed separately and might provide valuable information. However, path-based measures of (H)VGs are known to be strongly influenced by boundary effects~\cite{Donner2012AG}, so that they could possibly lose their discriminative power for {ir}reversibility tests.

\begin{acknowledgements}
This work has been financially supported by the Leibniz association (project ECONS) and the German National Academic Foundation (JFD). {(H)}VG analysis has been performed using the Python package \texttt{pyunicorn}. The EEG data in our real-world example have been kindly provided by the Department of Epileptology at the University of Bonn. We thank Yong Zou for {providing time series of} the Mackey-Glass equations.
\end{acknowledgements}

\bibliographystyle{eplbib}
\bibliography{reversibility_lib}

\end{document}